\documentclass[twocolumn]{svjour3}          % twocolumn
\smartqed  % flush right qed marks, e.g. at end of proof
\usepackage{graphicx}
\usepackage{xspace}
\usepackage{hyperref}
\usepackage{url}
\newcommand{\Openstack}{\texttt{Open\-Stack}\xspace}
\newcommand{\Roced}{{ROCED}\xspace}
\newcommand{\ATLAS}{ATLAS\xspace}
\newcommand{\CMS}{CMS\xspace}
\newcommand{\NEMO}{NEMO\xspace}
\newcommand{\Packer}{\texttt{Packer}\xspace}
\newcommand{\Puppet}{\texttt{Puppet}\xspace}
\newcommand{\Hieradata}{\texttt{Hieradata}\xspace}
\newcommand{\Moab}{\texttt{Moab}\xspace}
\newcommand{\Slurm}{\texttt{Slurm}\xspace}
\newcommand{\BeeGFS}{\textsc{BeeGFS}\xspace}
\newcommand{\OZ}{\texttt{Oz}\xspace}
\newcommand{\HTCondor}{\texttt{HTCondor}\xspace}
\newcommand{\kickstart}{\texttt{kickstart}\xspace}

\journalname{Computing and Software for Big Science}
\begin{document}

\title{Dynamic Virtualized Deployment of Particle Physics Environments on a
  High Performance Computing Cluster}
\author{Felix B\"uhrer \and Frank Fischer \and Georg Fleig \and Anton
  Gamel \and Manuel Giffels \and Thomas Hauth \and Michael Janczyk
  \and Konrad Meier \and
G\"unter Quast \and  Beno\^it Roland  \and Ulrike Schnoor \and
  Markus Schumacher \and Dirk von Suchodoletz \and Bernd Wiebelt
}

\institute{
   U. Schnoor \at   Universit\"at Freiburg, Physikalisches
  Institut, Hermann-Herder-Str. 3, 79104 Freiburg, Germany 
  \at \emph{Now at} CERN, CH-1211 Geneva
  23, Switzerland        %  \\
  \\\email{ulrike.schnoor@cern.ch}
  \and F. B\"uhrer \and B. Roland \and M. Schumacher \at Universit\"at Freiburg, Physikalisches
  Institut, Hermann-Herder-Str. 3, 79104 Freiburg, Germany
  \and A. Gamel  \at  Universit\"at Freiburg, Physikalisches
  Institut, Hermann-Herder-Str. 3, 79104 Freiburg, Germany \at  Universit\"at Freiburg, Rechenzentrum, Hermann-Herder-Str.10, 79104
  Freiburg, Germany
  \and M. Janczyk 
  \and K. Meier   \and D. von Suchodoletz 
 \and B. Wiebelt \at Universit\"at Freiburg, Rechenzentrum,
 Hermann-Herder-Str. 10, 79104
 Freiburg, Germany
  \and F. Fischer \and G. Fleig   \and M. Giffels  \and T. Hauth\and G. Quast  \at Karls\-ruher Institut f\"ur Technologie, Institut f\"ur Experimentelle
  Teilchenphysik, Wolfgang-Gaede-Str. 1, 76131 Karls\-ruhe, Germany
}

\date{20 December 2018}
% The correct dates will be entered by the editor

\maketitle

\begin{abstract}
The \NEMO High Performance Computing Cluster at the University of
Frei\-burg has been made available to
researchers of the \ATLAS and \CMS experiments.
Users access the cluster from external machines connected to the
World-wide LHC Computing Grid (WLCG).
 This paper describes how the full software environment of the WLCG
 is provided in a virtual machine image. The interplay between the
 schedulers for \NEMO and for the external
 clusters is coordinated through the \Roced service.
A cloud computing infrastructure is deployed at \NEMO to orchestrate the
simultaneous usage by bare metal and virtualized jobs.
Through the setup, resources are provided to users in a transparent,
automatized, and
on-demand way. The performance of the virtualized environment has been
evaluated for particle physics applications.


\end{abstract}

\section{Introduction}
\label{intro}
Particle physics experiments at the Large Hadron Collider (LHC) need a
great quantity of computing resources for data processing, simulation, and analysis.
This demand will be growing with the upcoming High-Luminosity upgrade of the LHC~\cite{HLLHCcompneeds}.
To help fulfill this requirement, High Performance Computing (HPC) resources provided by research institutions
can be useful supplements to the existing World-wide LHC Computing
Grid (WLCG)~\cite{wlcg} resources
allocated by the collaborations.

This paper presents the concepts and implementation of providing a HPC resource, the
shared research cluster \NEMO~\cite{nemo} at the University of Freiburg, to ATLAS and CMS users accessing external clusters connected to the WLCG with the purpose of accommodating data production as well as
data analysis on the HPC host system. The HPC cluster \NEMO at
the University of Freiburg is deploying an \Openstack~\cite{Openstack} instance to handle the
virtual machines. The challenge is in provisioning, setup, scheduling, and decommissioning the virtual research environments (VRE) dynamically and according to demand. For this purpose, the schedulers on \NEMO and on the external resources are
connected through the \Roced service~\cite{ROCED}.

A VRE in the context of this paper is a complete software stack
as it would be installed on a compute cluster fitted to the demands of ATLAS or CMS workloads.

\section{Virtualization infrastructure}
\label{sec:openstack}

Hardware virtualization has become mainstream technology over the last decade as it allows
to host more than one operating system on a single server and to strictly
separate users of software environments.
Hardware and software stacks are decoupled, such that complete software
environments can be migrated across hardware boundaries.
While widespread in computer center
operation this technique is rarely applied in HPC.

\subsection{Computing at the University of Freiburg}

The computer center at the University of Freiburg provides
medium scaled research
infrastructures like cloud, storage, and especially HPC services adapted to the
needs of various scientific communities. Significant standardization
in hardware and software is necessary for the operation of compute systems comprised of
more than 1000 individual nodes combined with a small group of administrators.

The level of granularity of the software stack provided is not fine enough to
directly support the requirements of world-wide efforts like the
ATLAS or CMS experiments.
Therefore, novel approaches are necessary to ensure optimal use of the system and to open the cluster to as many different use-cases as
possible without increasing the operational effort.
Transferring expertise from the operation of the established local
private cloud, 
the use of \Openstack as a cloud platform has been identified
as a
suitable solution for \NEMO. This approach provides a user defined software
deployment in addition to the existing software module system.
The resulting challenges range from the automated creation of suitable
virtual machines to their on-demand deployment and scheduling.

\subsection{Research Cluster \NEMO}

The research cluster \NEMO is a cluster for 
research in the federal state of Baden-W\"urttemberg in the scientific fields of Elementary Particle Physics, Neuroscience and
Microsystems Engineering. Operation started on  August 1, 2016.
It currently consists of 900 nodes with 20 physical cores and 128\,GiB of RAM each.
Omni-Path~\cite{Omnipath} spans a high-speed, low-latency network of 100\,Gbit/s between nodes.
The parallel storage has
768\,TB of usable capacity and is based on \BeeGFS~\cite{BeeGFS}.

A pre-requirement to execute a VRE is the efficient
provisioning of data which has to cross institutional boundaries in the CMS use-case.
A signficant bandwidth is needed to transfer the input data into the VRE from the storage system at
the Karlsruhe Institute of Technology (KIT) and to store back the results. The
NEMO cluster is connected with two 40\,Gbit/s links to the main router of the
University of Freiburg which itself is linked to the network of
scientific institutions in Baden-W\"urttemberg, BelW\"u, at
100\,Gbit/s.

\subsection{Separation of software environments}

The file system of a VRE is a
disk image presented as a single file. From the computer center's perspective
this image is a ``black box'' requiring no involvement or efforts like
updates of the operating system or the provisioning of software packages of a
certain version. From the researcher's perspective the VRE is an individual
virtual node whose operating system, applications and configurations
as well as certain hardware-level parameters, e.g. CPU and RAM, can be
configured fully autonomously by the researcher within agreed upon limits.

To increase the flexibility in hosted software environments, the standard bare metal
operation of \NEMO is extended with an installation of \Openstack
components~\cite{hpc-symp:2016}.
The \NEMO cluster uses Adaptive's Workload Manager \Moab~\cite{Moab} as a
scheduler of compute jobs.
\Openstack as well can schedule virtual machines on the same nodes and
resources.
To avoid conflicts, it is necessary to define the master scheduler
which decides the job assignment to the worker nodes.
Both \Moab and \Openstack are
unaware that another scheduler exists within the cluster and there is
no API which enables them to  communicate with each other. Since the majority of users still use the
bare metal HPC cluster, \Moab is deployed as the primary scheduler. It allows for
detailed job description and offers sophisticated scheduling features like
fair-share, priority-based scheduling, detailed time limits,
etc. \Openstack 's task is to deploy the virtual machines, but \Moab will initially start the VRE
jobs and the VRE job will instruct \Openstack to start the virtual machine on the
reserved resources with the required flavor, i.e. the resource definition in \Openstack.

When a VRE job is submitted to the \NEMO cluster, \Moab  first calculates the
priority and the needed resources of the job and then inserts it into its queue.
When the job is in line for execution and the requested resources are available,
the job  runs a script which then starts the VRE on the selected node
within the resource boundaries. During the run-time of the VRE a monitoring
script regularly checks if the VRE is still running and terminates the job when
the VRE has ended.
When the job ends, \Openstack gets a signal to terminate the virtual machine and
the VRE job ends as well.  Neither \Moab nor \Openstack have access
inside the VRE, so they cannot assess if the VRE is actually busy or
idle.
The software package \Roced (described in
further detail in Section~\ref{section:roced}) has been introduced to
solve this issue.
It is used as a broker between
different HPC schedulers,  translating resources and monitoring usage inside the
virtual machine, as well as starting and stopping VRE images on demand.

\section{Generation of the VRE image}
The VREs for ATLAS and CMS software environments consist in \Openstack containers
in the format of compatible VM images.
These images are provided in an automatized
way allowing versioning and archiving of the environments captured in
the images.

\subsection{\Packer combined with \Puppet}

One approach to generate the image is the open-source tool
\Packer~\cite{packer}, interfaced to the system configuration framework \Puppet~\cite{puppet}.
\Packer allows to configure an image based on an ISO image file using a \kickstart~\cite{kickstart} file and flexible script-based configuration. 
It also provides an interface to \Puppet making it particularly convenient if an existing \Puppet role is to be used for the images. If the roles are defined according to the hostname of the machine as is conventional in \Puppet with \Hieradata, the hostname needs to be set in the scripts supplied to \Packer. Propagation of certificates requires an initial manual start of a machine with the same hostname to allow handshake signing of the certificate from the \Puppet server.

\Packer's interface to \Puppet allows a fully automated image generation with up-to-date and version-controlled configuration. At the end of the generation run, the image is  transferred to the \Openstack image server.

\subsection{Image generation using the \OZ toolkit}

Another option to employ a fully-automated procedure is to use the \OZ toolkit~\cite{OZ}. All requirements and configuration options of an image can be specified through a XML template file. The partitioning and installation process of the operating system is fully automated, as \OZ will use the remote-control capabilities of the local hypervisor. After the installation of the operating system, additional libraries and configuration files can be installed. Once the image has been created, it is automatically compressed and uploaded to a remote cloud site.
This technique allows to build images in a reproducible fashion, as all templated files are version controlled using \texttt{git}. Furthermore, existing template files are easy to adapt to new sites and experiment configurations.

\section{Interfacing batch systems and virtual resources using \Roced}
\label{section:roced}
While HPC systems with support for virtualized research environments and commercial cloud providers offer the
necessities to acquire computing and storage capacity by dynamic
resource booking, the computing needs of high energy physics
re\-search groups ad\-di\-tion\-al\-ly require work\-flow
ma\-na\-ge\-ment sys\-tems capable of maintaining thousands of batch
jobs. Some cloud providers, for example Amazon with AWS
Batch~\cite{awsbatch}, provide a service for workflow management,
however these offers are often limited to one specific cloud instance. To dynamically distribute batch jobs to multiple sites and manage machine life-time on specific sites, a combination of a highly-scalable batch system and a virtual machine scheduler is desirable.

\subsection{\Roced}
\begin{figure*}
\begin{center}
  \includegraphics[width=0.9\linewidth]{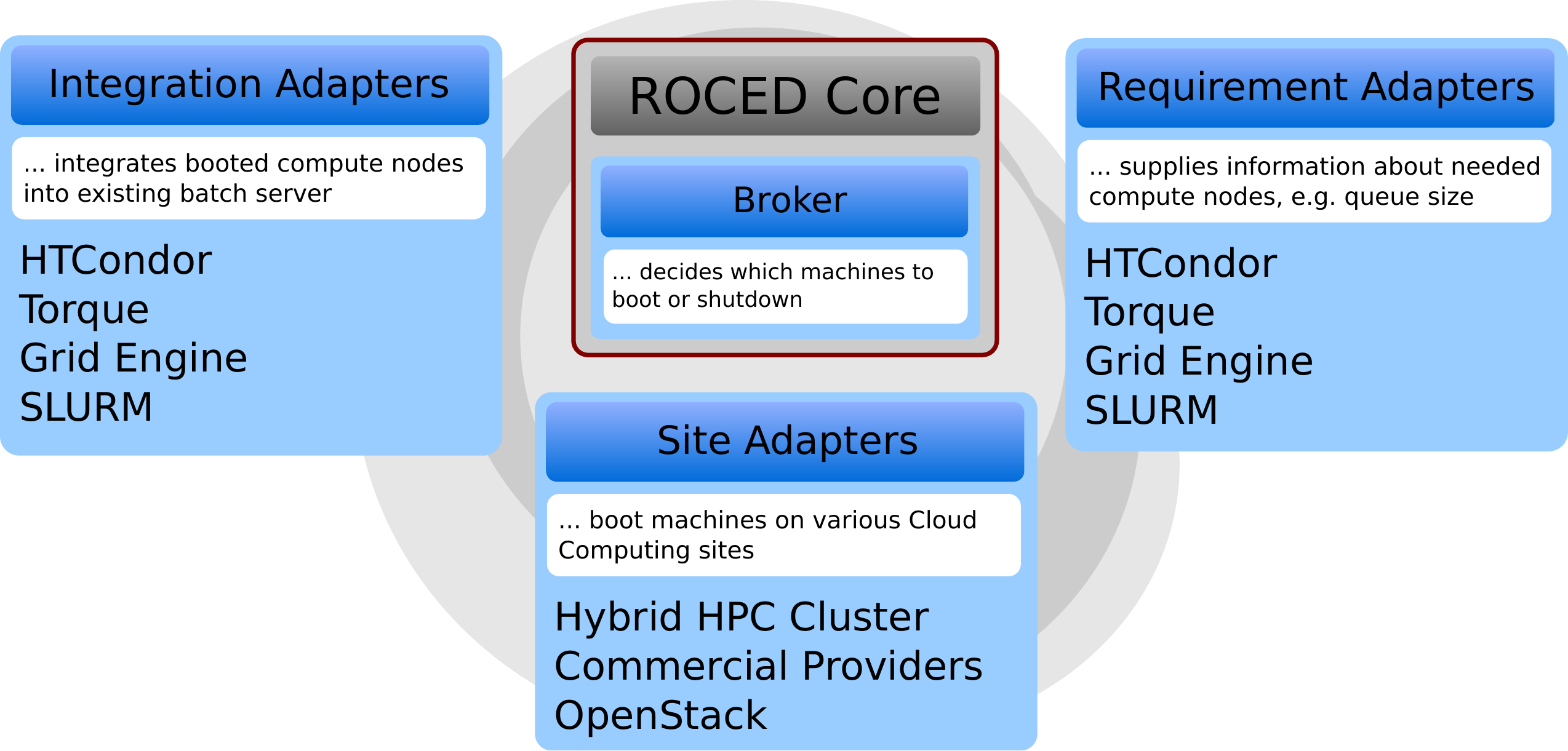}
  \caption{Overview of the \Roced modular design. The  \Roced Core contains the Broker which decides when and on which sites new virtual machines are booted. The Requirement Adapters report about the utilization and resource requirements of the attached batch systems. The Site Adapter is responsible to manage the lifetime of virtual machines on a cloud site and the Integration Adapters ensure that newly booted machines are integrated into the batch system.}
  \label{fig-roced}
\end{center}
\end{figure*}

Many capable batch systems exist today and they can be interfaced to virtualization providers using the cloud meta-scheduler \Roced (Responsive On-demand Cloud Enabled Deployment) which has been developed at the KIT since 2010~\cite{ROCED}. \Roced is written in a modular
fashion in python and the interfaces to batch systems and cloud sites
are implemented as so-called \textit{Adapters}. This makes \Roced
independent of specific user groups or workflows. It provides a
scheduling core which collects the current requirement of computing
resources and decides if virtual machines need to be started or can be
stopped. One or more Requirement Adapters report the current queue
status of batch systems to the central scheduling core. Currently,
Requirement Adapters are implemented for the Slurm, Torque/Moab, HTCondor
and GridEngine batch systems. The Site Adapters allow \Roced to start,
stop, and monitor virtual machines on multiple cloud
sites. Implementations exist for Amazon EC2, OpenStack, OpenNebula and
Moab-based virtualization at HPC centers. Special care has been put
into the resilience of \Roced: it can automatically terminate
non-responsive machines and restart virtual machines in case some
machines have dropped out. This allows VM setups orchestrated by \Roced with thousands of virtual machines and many tens of thousands of jobs to run in production environments.
The modular design of \Roced is shown in Fig.~\ref{fig-roced}.

\subsection{Using \HTCondor as front-end scheduler}\label{sec:ROCED:HTCondor}
The open-source project \HTCondor provides a workload management system which is highly configurable and modular~\cite{HTCondor}. Batch processing workflows can be submitted and are then forwarded by \HTCondor to idle resources. \HTCondor maintains a resource pool, which worker nodes in a local or remote cluster can join. Once \HTCondor has verified the authenticity and features of the newly joined machines, computing jobs are automatically transferred. Special features are available to connect from within isolated network zones, e.g. via a Network Address Translation Portal, to the central \HTCondor pool. The Connection Brokering (CCB) service~\cite{HTCondorCCB} is especially valuable to connect virtual machines to the central pool. These features and the well-known ability of \HTCondor to scale to O(100k) of parallel batch jobs makes \HTCondor well suited as a workload management system for the use cases described in this paper.

The CMS group at the KIT is using \HTCondor for scheduling the jobs to be submitted to \NEMO. The VRE for CMS contains the \HTCondor client \texttt{startd}.
This client is started after the machine has fully booted and connects to the central \HTCondor pool at KIT via a shared secret. Due to \HTCondor's dynamic design, new machines in the pool will automatically receive jobs and the transfer of the job configuration and meta-data files is handled via \HTCondor's internal file transfer systems.

\subsection{Using \Slurm as front-end scheduler}

Alternatively to the approach described above, the
open-source workload managing system \Slurm~\cite{Slurm} has been interfaced into \Roced by
the ATLAS group at University of Freiburg.
\Slurm provides a built-in functionality for the dynamic
startup of resources in the \textit{Slurm Elastic Computing}
module~\cite{SlurmElastic}.
However, this module is based on the assumption of a fixed maximum
startup time of the machines.
In the considered case, due to the queue in the host system, the start of a
resource can be delayed by a significant, varying time period.
In addition the transfer of information, such as error states, from one scheduler to the
other, and therefore to the user, is  limited.
Therefore, \Roced has been chosen as the interface between the
\Moab scheduler on the host system and the \Slurm
scheduler on the submission side.

\begin{figure}

\includegraphics[width=0.95\linewidth]{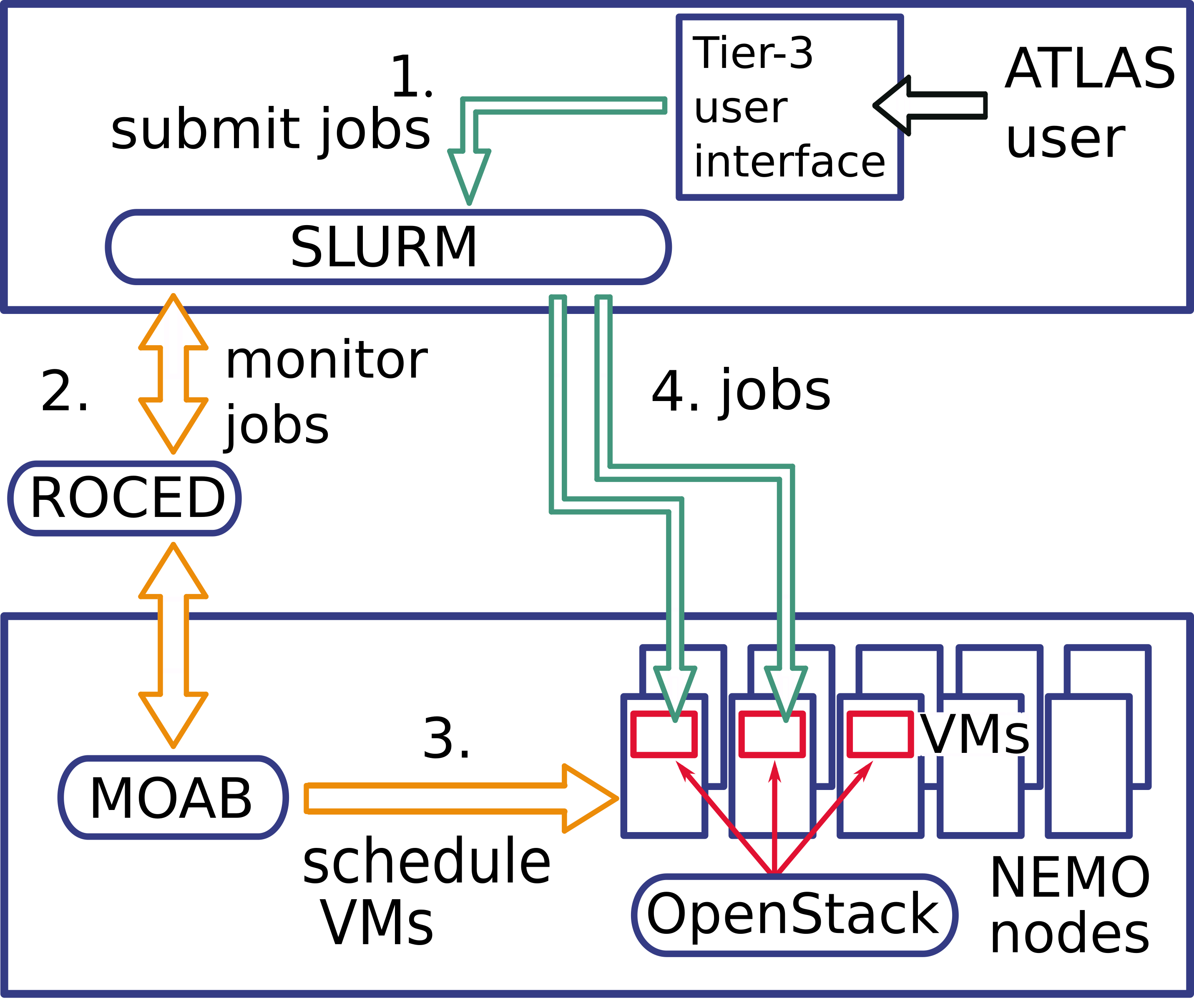}
\caption{Implementation of \Roced with \Slurm on the Tier-3 cluster of
  the WLCG used by ATLAS researchers in Freiburg.}
\label{fig:slurmRocedBFG}
\end{figure}

The scheduling system is illustrated in Fig.~\ref{fig:slurmRocedBFG}.
For \Slurm, it is necessary that each potential virtual
machine is registered in the configuration at the time of start of the
\Slurm server as well as the client. \Slurm configurations also
need to be in agreement between server and client.
Therefore, a range of hostnames is registered in the configuration in
a way that is mapped to potential IP addresses of virtual machines.
These virtual machines have a fixed number of CPUs and memory assigned and are
registered under a certain \Slurm partition.
When a job is submitted to this partition and no other resource is
available, information from the \Slurm \texttt{squeue} and
\texttt{sinfo} commands is requested and parsed for the required information.

Since the ATLAS Freiburg group comprises three sub-groups, each mapped
to a different production account on \NEMO, special care is taken to
avoid interference of resources used by another account to ensure fair share on \NEMO, while
nevertheless allowing jobs from one group to occupy otherwise idle resources of another group.

\Roced determines the amount of virtual machines to be started and sends the
corresponding VRE job submission commands to \Moab.
After the virtual machine has booted, the hostname is set to the IP
dependent name which is known to the \Slurm configuration.
A \texttt{cron} job executes several sanity
checks on the system.
Upon successful execution of these tests, the \Slurm client
running in the VM starts accepting the queued jobs.
After completion of the jobs and a certain period of receiving no new jobs from the queue, the
\Slurm client in the machine drains itself and the machine
shuts itself down.
The IP address as well as the corresponding hostname in \Slurm
are released and can be reused by future VREs.

\section{Analysis of performance and usage}

The ROCED-based solution described above has been implemented and put into production by the research groups at the University of Freiburg (Institute of Physics) and the KIT (Institute of Experimental
Particle Physics). To prove the usefulness of this approach
statistical analyses of the performance of the virtualized setup both
in terms of CPU benchmarks and usage statistics have been conducted.

\subsection{Benchmarks}
Benchmark tests are performed with the primary goal to measure the performance of the CPU for High Energy Physics applications. 
Alongside the legacy HEP-SPEC06 (HS06) benchmark \cite{Hepspec}, the performance of the compute resources is furthermore evaluated with the ATLAS Kit Validation
KV \cite{DeSalvo:2010zza}, a fast benchmark developed to provide real-time information of the WLCG performance and available in the CERN benchmark suite \cite{Alef:2017jyx}.
The primary target is to measure the performance of the CPU for High Energy Physics applications.
The KV benchmark is making use of the simulation toolkit GEANT4 \cite{Agostinelli:2002hh} to simulate the interactions of single muon events in the detector of the ATLAS experiment
and provides as ouput the number of events produced per second. It constitutes a realistic workload for High Energy Physics jobs. \\

To assess the impact of the virtualization, the performance of the identical hardware configuration (20 cores Intel Xeon E5-2630 CPUs) has been determined either deployed via
the standard bare metal operation on the \NEMO cluster (\NEMO bare metal) and on the ATLAS Tier-3 center in Freiburg (ATLAS Tier-3 bare metal), or as virtual machines on the
\NEMO cluster (\NEMO VM). On the ATLAS Tier-3 bare metal and on the virtual machines running on the \NEMO cluster, hyper-threading (HT) technology is activated. Both are using Scientific
Linux 6 \cite{SL6} as the operating system.
On the cluster \NEMO bare metal jobs are restricted to 20 cores by cgroups, since the application mix is broader than on HEP clusters. The operating system is CentOS7 \cite{CentOS7}.
The scores of the HEP-SPEC06 and KV benchmarks have been determined for these three configurations as a function of the number of cores actually used by the benchmarking processes.
This number ranges from 2 to 40 for the ATLAS Tier-3 bare metal and for the \NEMO VM, for which HT is enabled, and from 2 to 20 for the \NEMO bare metal, for which HT is not implemented.
The benchmarks have been run 20 times for each core multiplicity value, and the means and standard deviations of the corresponding distributions have been extracted. \\

\begin{figure}[htbp]
\centering
\includegraphics[width=0.44\textwidth]{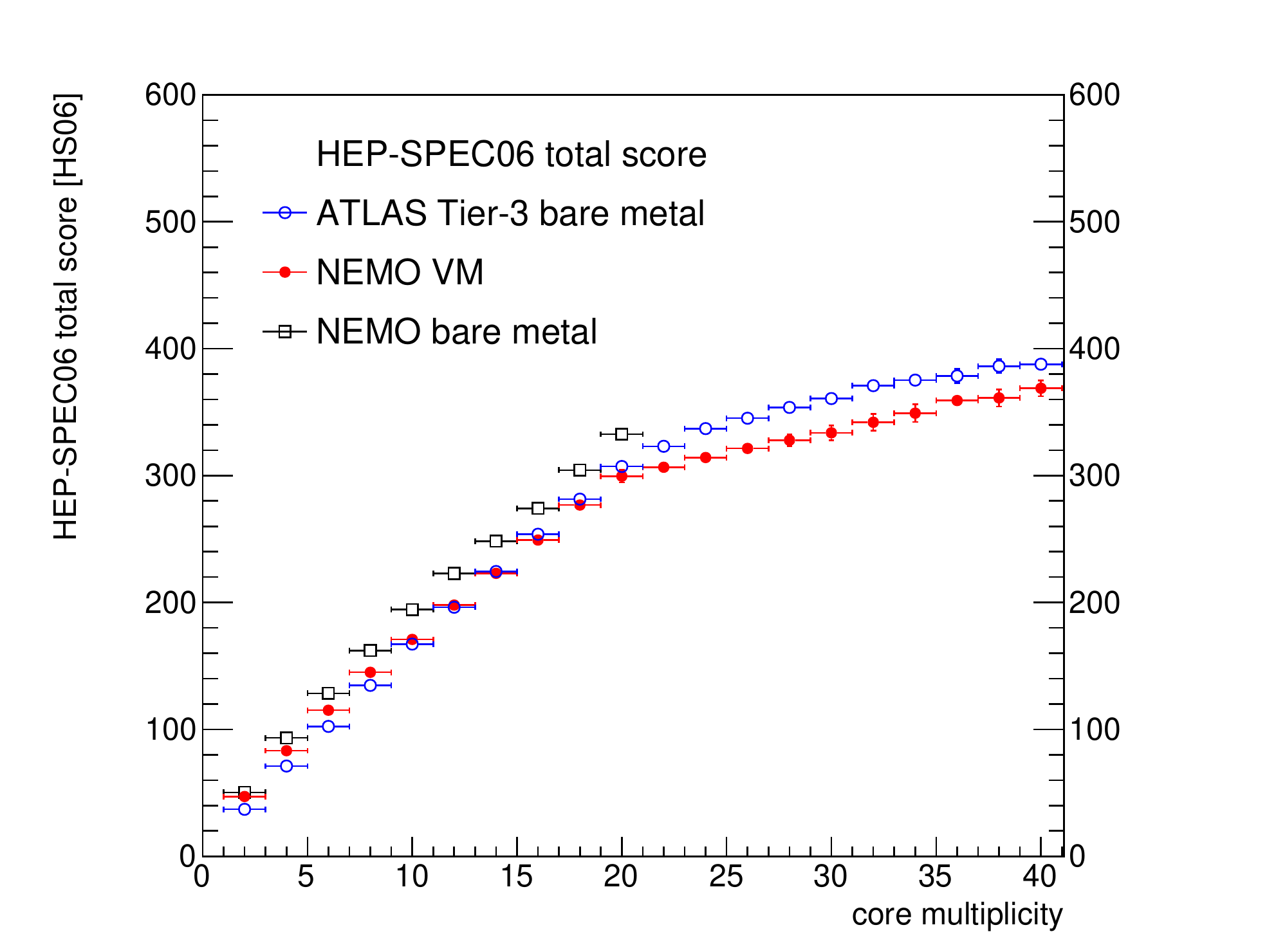}
\includegraphics[width=0.44\textwidth]{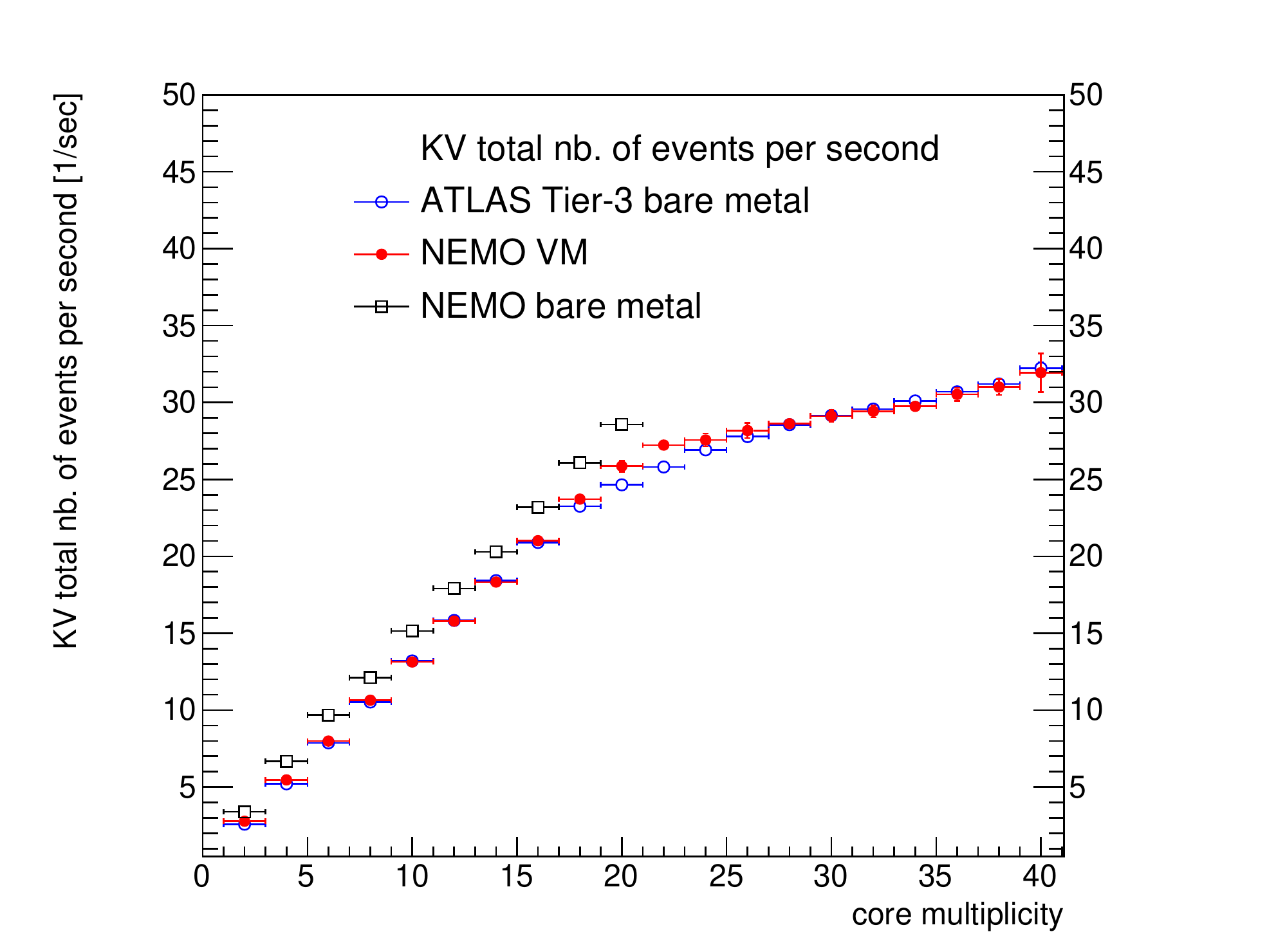} 
\caption{Total score as a function of the core multiplicity for the HEP-SPEC06 (top) and KV (bottom) benchmarks for the ATLAS Tier-3 bare metal (blue open circles),
the \NEMO VMs (red full circles) and the \NEMO bare metal (black open squares). The data points represent the average values of the benchmarks for each core multiplicity,
and the vertical bars show the associated standard deviations.}
\label{bmk-total}
\end{figure}

The HEP-SPEC06 and KV results are presented in Figure \ref{bmk-total} for the three configurations considered.
The total scores of the two benchmarks are increasing until the maximum number of physical cores has been reached, and are characterized by a flattening increase afterwards.
The scores of the virtual machines running on the \NEMO cluster are only slightly lower than those obtained for the \NEMO bare metal, and the loss of performance
due to the virtualization does not exceed 10$\%$.
For the VMs running on the \NEMO cluster and the ATLAS Tier-3 bare metal, the interplay between the virtualization and the different operating systems leads to very similar scores
for the two configurations, particularly for the KV benchmark, and the loss of performance is smaller than 10$\%$ as well.

\subsection{Usage statistics}
\begin{figure}
\begin{center}
  \includegraphics[width=1.0\linewidth]{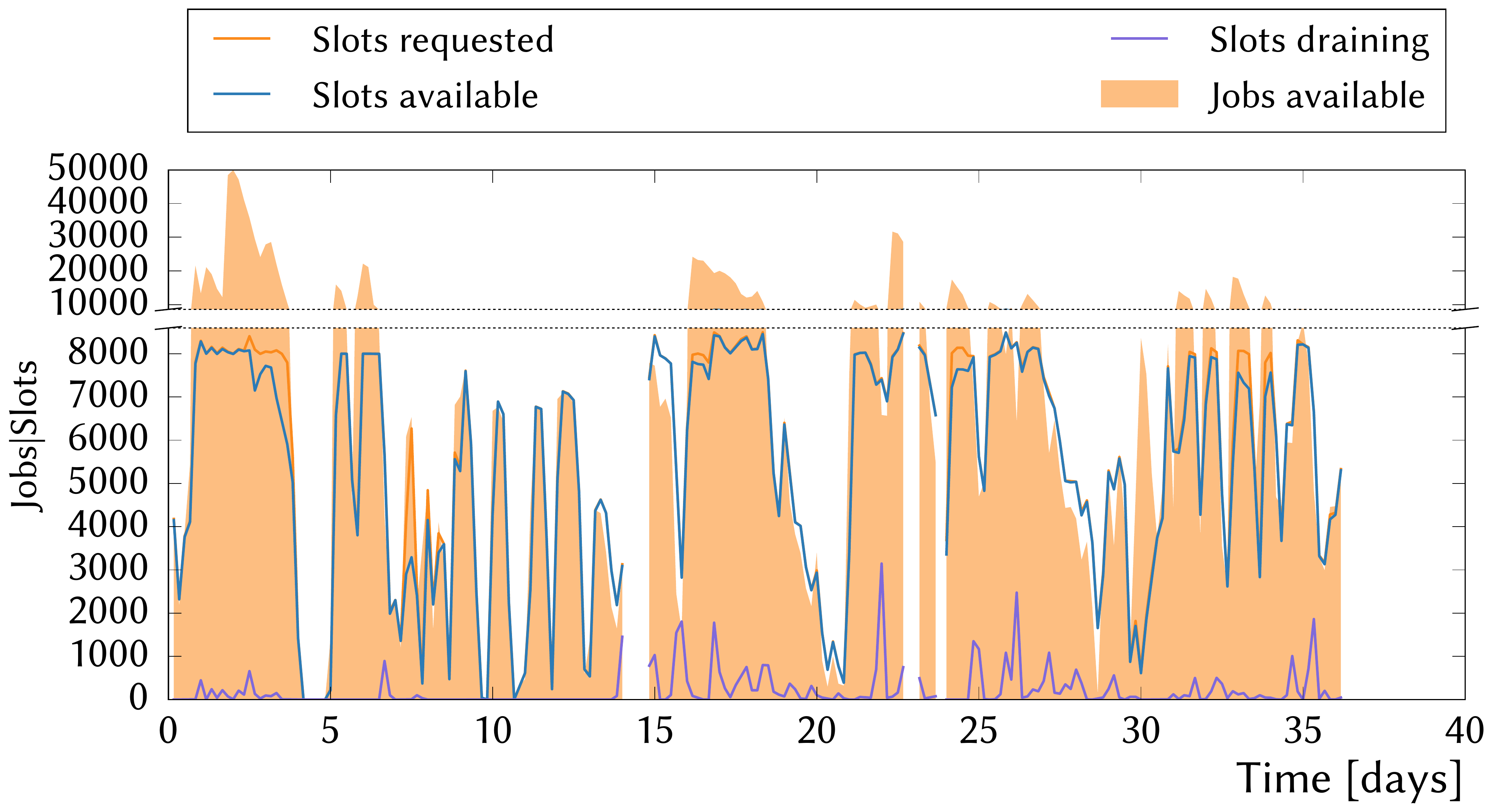}
  \caption{Utilization of the shared HPC system by booted virtual machines. Up to 9000 virtual cores were in use at peak times. The fluctuations in the utilization reflects the patterns of the submission of jobs by the CMS users at the physics institute in Karlsruhe. The number of draining slots displays the amount of job slots still processing jobs while the rest of the node's slot are already empty.}
  \label{fig-frplots}
\end{center}
\end{figure}

Fig. \ref{fig-frplots} shows the utilization of virtual machines which were orchestrated by \Roced depending on the resource demands of the users of the KIT group.
At peak times, up to 9000 virtual cores were filled with user jobs, consuming more than a half of the initial 16000 \NEMO cores.

\begin{figure}
\begin{center}
  \includegraphics[width=\linewidth]{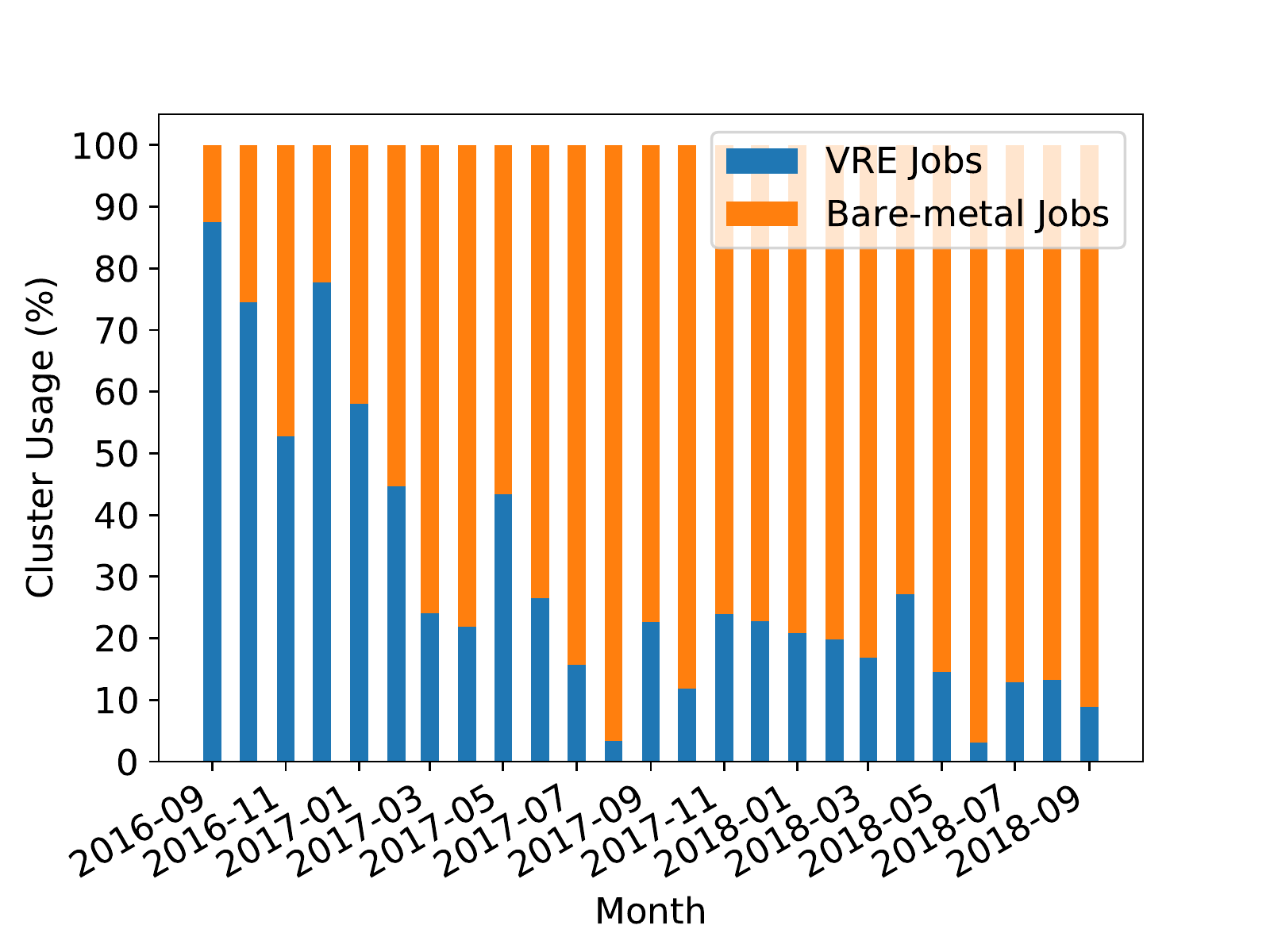}
  \caption{Estimated usage of the \NEMO cluster in the time from September 2016
    to September 2018. The orange bars indicate the usage by jobs
    running directly in the hosts' operating system, while the blue bars are jobs
    running in virtual machines. The decrease of VRE jobs is partially explained
    by an increasing number of bare metal jobs submitted.}
  \label{fig-nodeusage}
\end{center}
\end{figure}

The usage of the hybrid cluster model is presented in Fig.~\ref{fig-nodeusage}.
The diagram shows the shared usage of \NEMO's cluster nodes running either
bare-metal or virtualized jobs. The part of the cluster which runs virtualized
jobs changes dynamically from job to job, since the VREs are started by
a standard bare-metal job.

At the beginning the cluster was only containing the operating system and some
basic development tools. Scientific software was added after the cluster was
already in production mode. Since the VRE for the CMS project was already
available when the \NEMO cluster started, it could already use the whole cluster
while other groups still had to migrate from other ressources. This explains the high usage by VREs in the first months of
operation. With more and more software being available for bare-metal usage the
fraction of VRE jobs decreased.

\section{Conclusions and Outlook}

A novel system for the dynamic, on-demand provisioning of virtual machines to run jobs
in a high energy physics context on an external, not dedicated resource as
realized at the HPC cluster \NEMO at the University of Freiburg has been
implemented. An interface between the schedulers of the
host system and the external systems from which requests are sent is needed to
monitor and steer jobs in a scalable way. For this workflow the cloud meta-scheduler \Roced
has been implemented and deployed for the described use-cases.
The approach can be adapted to work with other platforms and could be extended to
container technologies like Singularity~\cite{VRE2017}.

The CPU performance and usage of the setup have been analyzed for the job execution environment.
The expected performance loss due to the virtualization has been found to be
sufficiently small to be compensated by the added flexibility and other benefits
of this setup.

A possible extension of such a virtualized setup is the provisioning of functionalities
for snapshots and migration of jobs. This would facilitate the efficient integration of
long-running monolithic jobs into HPC clusters.

The provided solution extends the available compute ressources for HEP calculations
and could be one possibilitly to cope with new data from the upcoming High-Luminosity
upgrade of the LHC. Since HEP VREs are perfect for backfilling this could be used on various
cluster ressources.

%\subsection{Subsection title}
%\label{sec:2}
%as required. Don't forget to give each section
%and subsection a unique label (see Sect.~\ref{sec:1}).

%\paragraph{Paragraph headings} Use paragraph headings as needed.
%\begin{equation}
%a^2+b^2=c^2
%\end{equation}
%
%% For one-column wide figures use
%\begin{figure}
%% Use the relevant command to insert your figure file.
%% For example, with the graphicx package use
%  \includegraphics{example.eps}
%% figure caption is below the figure
%\caption{Please write your figure caption here}
%\label{fig:1}       % Give a unique label
%\end{figure}
%%
%% For two-column wide figures use
%\begin{figure*}
%% Use the relevant command to insert your figure file.
%% For example, with the graphicx package use
%  \includegraphics[width=0.75\textwidth]{example.eps}
%% figure caption is below the figure
%\caption{Please write your figure caption here}
%\label{fig:2}       % Give a unique label
%\end{figure*}
%
%% For tables use
%\begin{table}
%% table caption is above the table
%\caption{Please write your table caption here}
%\label{tab:1}       % Give a unique label
%% For LaTeX tables use
%\begin{tabular}{lll}
%\hline\noalign{\smallskip}
%first & second & third  \\
%\noalign{\smallskip}\hline\noalign{\smallskip}
%number & number & number \\
%number & number & number \\
%\noalign{\smallskip}\hline
%\end{tabular}
%\end{table}

\begin{acknowledgements}
This research is supported by the Ministry of Science, Research and the Arts Baden-W\"urttemberg through the bwHPC grant
and by the German Research Foundation (DFG) through grant no INST
39/963-1 FUGG for the bwForCluster NEMO.
The work of F.B. and T.H. was supported by the Virtual Open Science
Collaboration Environment (ViCE) project MWK 34-7547.221 funded by the
Ministry of Science, Research and the Arts Baden-W\"urttemberg.
The work of U.S. was supported by  the Federal Ministry of Education
and Research within the project 05H15VFCA1
``Higgs-Physik mit dem und Grid-Computing f\"ur das ATLAS-Experiment
am LHC''.
The work of T.H. was supported by the Federal Ministry of Education
and Research within the project 05H15VKCCA ``Physik bei h\"ochsten Energien mit dem CMS-Experiment
    am LHC''.
\end{acknowledgements}

% BibTeX users please use one of
%\bibliographystyle{spbasic}      % basic style, author-year citations
%\bibliographystyle{spmpsci}      % mathematics and physical sciences
%\bibliographystyle{spphys}       % APS-like style for physics
%\bibliography{}   % name your BibTeX data base

% Non-BibTeX users please use

\end{document}